# Capillary stamping of functional materials: parallel additive substrate patterning without ink depletion

Mercedes Runge, Hanna Hübner, Alexander Grimm, Gririraj Manoharan, René Wieczorek, Michael Philippi, Wolfgang Harneit, Carola Meyer, Dirk Enke, Markus Gallei, Martin Steinhart[*]

Dr. Mercedes Runge, Michael Philippi, Prof. Martin Steinhart

Institute of Chemistry of New Materials and CellNanOs, University of Osnabrück, 49069 Osnabrück, Germany

E-mail: martin.steinhart@uos.de

Hanna Hübner, Prof. Markus Gallei

Chair in Polymer Chemistry, Saarland University, 66123 Saarbrücken, Germany

Gririraj Manoharan, René Wieczorek, Prof. Wolfgang Harneit, Dr. Carola Meyer

Physics department, University of Osnabrück, 49069 Osnabrück, Germany

Alexander Grimm, Prof. Dirk Enke

Institute of Chemical Technology, Universität Leipzig, 04103 Leipzig, Germany

ABSTRACT

Patterned substrates for optics, electronics, sensing, lab-on-chip technologies, bioanalytics, clinical diagnostics as well as translational and personalized medicine are typically prepared by additive substrate manufacturing including ballistic printing and microcontact printing. However, ballistic printing (e.g., ink jet and aerosol jet printing, laser-induced forward transfer) involves serial pixel-by-pixel ink deposition. Parallel additive pattering by microcontact printing is performed with solid elastomeric stamps suffering from ink depletion after a few stamp-substrate contacts. The throughput limitations of additive state-of-the art patterning thus arising may be overcome by capillary stamping – parallel additive substrate patterning without ink depletion by mesoporous silica stamps, which enable ink supply through the mesopores anytime during stamping. Thus, either arrays of substrate-bound nanoparticles or colloidal nanodispersions of detached nanoparticles are accessible. We processed three types of model inks: 1) drug solutions, 2) solutions containing metallopolymers and block copolymers as well as 3) nanodiamond suspensions representing colloidal nanoparticle inks. Thus, we obtained aqueous colloidal nanodispersions of stamped drug nanoparticles, regularly arranged ceramic nanoparticles by post-stamping pyrolysis of stamped metallopolymeric precursor nanoparticles and regularly arranged nanodiamond nanoaggregates. Capillary stamping may overcome the throughput limitations of state-of-the-art additive substrate manufacturing while a broad range of different inks can be processed.

**Keywords:** substrate manufacturing, lithography, microcontact printing, porous materials, nanoparticles

# 1. Introduction

Substrate manufacturing[1] is relevant to application fields like optics, electronics, sensing, lab-on-chip technologies, bioanalytics, clinical diagnostics, translational medicine and personalized medicine. An important embodiment of substrate manufacturing is the additive lithographic deposition of functional inks; in this way, substrates may be patterned with functional materials and tailored nanoparticle arrays may be generated. An ideal additive lithographic substrate patterning technique would allow the execution of an unlimited number of parallel large-area ink deposition steps characterized by short cycle times without process-intrinsic interruptions caused, for example, by ink depletion. However, state-of-the-art substrate patterning techniques do not meet this requirement. Contactless ballistic pattering methods including inkjet printing, aerosol jet printing[2] and laser-induced forward transfer[3] can, in principle, be carried out continuously because ink can continuously be supplied to the substrate. However, these methods involve serial pixel-by-pixel writing associated with limitations regarding patternable areas and/or throughput. Additive substrate patterning by microcontact printing[4] and variations thereof such as polymer pen lithography (PPL),[5] capillary force lithography,[6] wet lithography[7] and particle transfer printing[8] are parallel and allow simultaneous patterning of large substrate areas. However, these methods involve transfer of ink coated on the outer surfaces of solid elastomeric stamps to substrates. Consequently, ink depletion in the course of successive stamp-substrate contacts results in deteriorating quality of the stamped patterns. Automated stamping devices for parallel additive surface manufacturing would be commercially available,[9] but stamping needs to be interrupted after a limited number of stamping steps to recoat the solid elastomeric stamps with ink. Moreover, nonambient conditions to prevent the stamps from drying out need to be applied.

Capillary stamping with stamps penetrated by continuous spongy pore systems[10] open to the environment is a parallel substrate patterning method like classical microcontact printing that nevertheless overcomes the problem of ink depletion. Similar to the above-mentioned ballistic patterning methods, ink can be supplied continuously. A porous stamp is soaked with ink (Figure 1a) and approached to a substrate until the ink forms capillary bridges between the stamp's contact elements and the substrate (Figure 1b). The capillary bridges are hydraulically coupled to the ink located in the stamp's pore system, which is an ink reservoir that can be backfilled anytime. Solvent evaporation drives the nonvolatile ink components into the capillary bridges.[10f, 11] When the porous stamp is retracted, the ink capillary bridges rupture. At the positions of the capillary bridges, ink droplets, in which the nonvolatile ink components are enriched, remain on the substrate. Solvent evaporation and optional post-stamping treatments convert the ink droplets into nanoparticles of a functional target material (Figure 1c). Since ink can be supplied to the contact surface of the stamp via the stamp's pore system anytime, an unlimited number of successive stamping steps can be performed without interruptions for reinking.

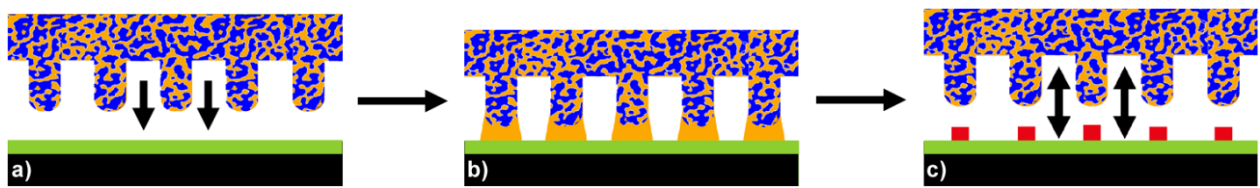

**Figure 1.** Schematic illustration of capillary stamping. a) A spongy mesoporous silica stamp (blue) filled with ink (orange) is approached to a substrate (black) optionally modified with a coating (green), for example, perfluorinated silanes. Once the distance is small enough, ink capillary bridges form between the contact elements of the mesoporous silica stamp and the substrate. c) Nonvolatile ink components are deposited on the substrate at the positions of the capillary bridges and form nanoparticles (red) after the retraction of the mesoporous silica stamp.

Here we evaluate capillary stamping with mesoporous silica stamps as parallel additive substrate patterning technique that potentially overcomes the drawbacks of the state of the art. For this purpose, model inks representing three important ink categories were tested: 1) inks containing organic low molecular mass compounds such as active pharmaceutical ingredients (APIs), 2) inks containing polymers and 3) colloidal nanoparticle inks. Solutions containing the synthetic estrogen 17α-ethinylestradiol (EE$_2$) were tested as examples for inks containing functional low-molecular mass compounds because additive manufacturing is considered as a promising access to drug delivery and testing systems.[12] Moreover, we detached the stamped EE$_2$ nanoparticles from the substrates and transferred them to aqueous colloidal nanodispersions. Colloidal nanodispersions containing nanoparticles of poorly water-soluble APIs like EE$_2$ are potentially advantageous delivery systems for oral, dermal, ocular, parenteral and pulmonary administration routes,[13] because API nanoparticles have large specific surfaces areas enhancing solubility and bioavailability as compared to conventional dosage forms.[14]

As examples for inks containing polymers, which have attracted significant interest for decades,[15] solutions containing poly(dimethylferrocenylsilane) (PFS) homopolymer or the block copolymers (BCPs) poly(ferrocenyldimethylsilane)-*block*-poly(2-vinylpyridine) (PFS-*b*-P2VP; Supporting Figure S1)[16] and polystyrene-*block*-poly(2-vinylpyridine) (PS-*b*-P2VP)[17] were tested. Pyrolysis of PFS yields ceramic pyrolysis products;[18] pyrolysis of PFS patterns generated by confined evaporation of PFS solutions[19] and of PFS nanorods confined to nanoporous anodic alumina[20] yielded the corresponding nano- and micropatterned ceramics. Common applications of PFS-*b*-P2VP include BCP lithography,[21] as well as the growth of carbon nanotubes using its pyrolysis products as catalysts.[22] We show that pyrolysis of PFS and PFS-*b*-P2VP nanoparticles generated by capillary stamping yields ordered arrays of Fe/SiC ceramic nanostructures. PS-*b*-P2VP solutions were tested as inks to study how stamped PS-*b*-P2VP patterns can be further modified by post-stamping solvent treatments triggering solvent-induced morphology reconstruction.[23] Colloidal inks containing nanoparticles are, for example, applied in the field of printed electronics.[24] In a previous work, stripes with widths in the 10 μm range consisting of 60 nm silica nanoparticles were stamped with porous polymer stamps obtained by a phase separation route having pores with diameters of several 100 nm.[10a] Here, we tested whether colloidal nanodispersions of nanodiamond (ND) nanoparticles are processable using mesoporous silica stamps, aiming at additive lithographic generation of small ND aggregates. NDs were selected because of their broad application range including cell tracking, super-resolution imaging and nanoscale temperature sensing,[25] as well as their use as cellular biomarkers.[26] So far, the lithographic deposition of small ND aggregates has remained challenging.[27]

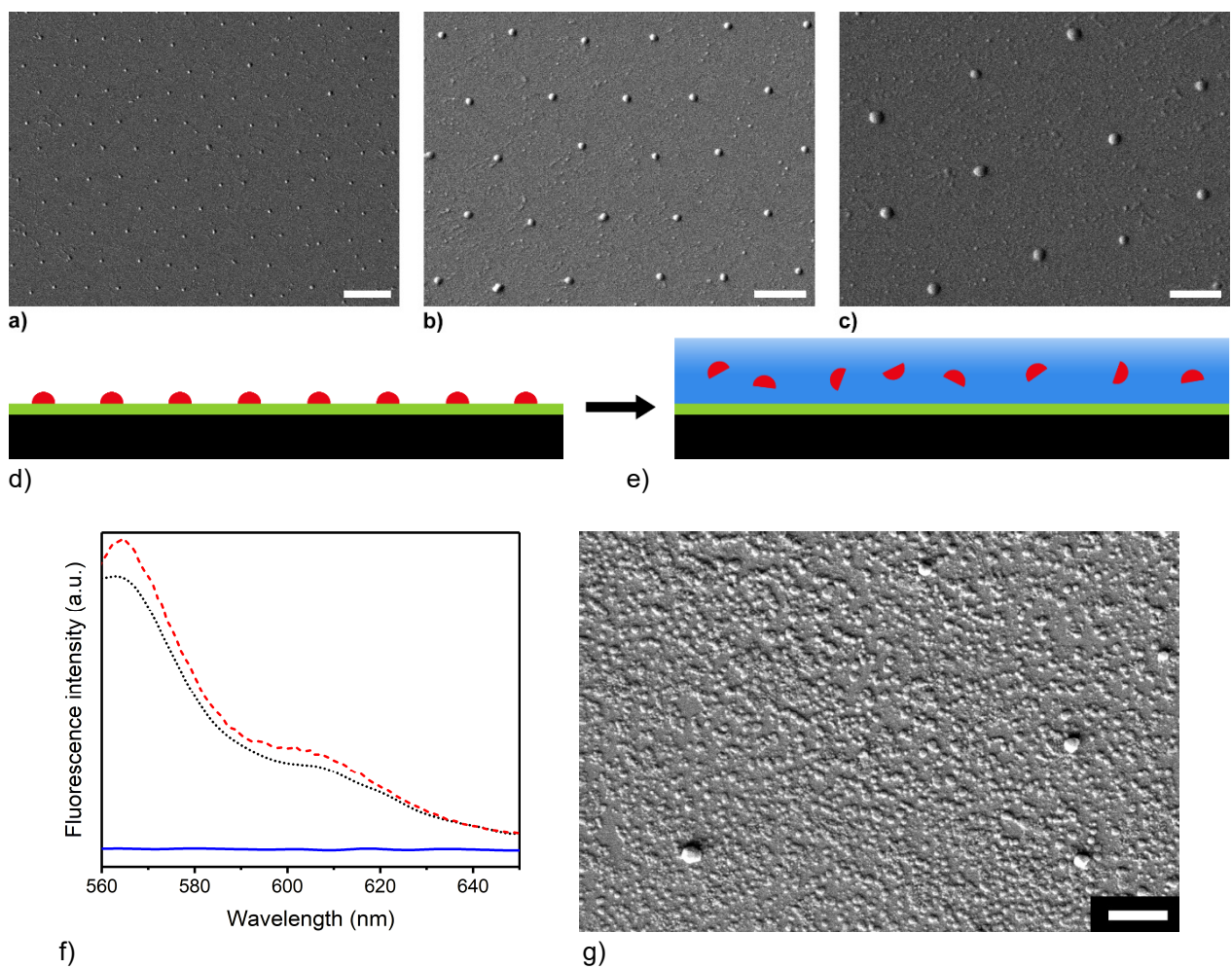

**Figure 2.** Capillary stamping of 17α-ethinylestradiol (EE$_2$) nanoparticles. a)-c) SEM images of EE$_2$ nanoparticle arrays stamped on FDTS-modified glass slides. a) Overview; b), c) details. The lengths of the scale bars correspond to 2 µm (panel a), 1 µm (panel b) and 600 nm (panel c). d), e) Schematic diagram of the transfer of EE$_2$ nanoparticles (red) from glass slides (black) covered with FDTS (green) to aqueous colloidal nanodispersions (blue). f) Fluorescence spectra (excitation wavelength 540 nm) of EE$_2$ nanoparticles dispersed in water prior to (blue solid line) and after (black dotted line) click-reaction with sulfo-cyanine-3-azide (excess sulfo-cyanine-3-azide was removed by dialysis). The dashed red curve is a fluorescence spectrum of a 1 mM sulfo-cyanine-3-azide solution in deionized water (excitation wavelength 530 nm). g) SEM image of EE$_2$ nanoparticles prepared by capillary stamping that were transferred to a colloidal nanodispersion and then deposited onto a glass slide. The scale bar corresponds to 1 µm.

## 3. Results and discussion

### 3.1 Capillary stamping with mesoporous silica stamps as process platform

To test capillary stamping for parallel additive surface patterning, we selected mesopororous silica stamps[10f, 10g] synthesized with methyltrimethoxysilane (MTMS) as trifunctional silica precursor[28] instead of polymeric porous stamps predominantly used so far [10a-e] for the following reasons. 1) The mesoporous silica stamps exhibit outstanding inertness against organic solvents – organic solvents do not swell, deform or dissolve silica. In contrast, polymeric porous stamps are characterized by limited compatibility with organic solvents as organic solvents may swell, deform or dissolve polymers. 2) The mesoporous silica stamps used here were derived from the trifunctional silica precursor MTMS. As a result, they exhibit non-polar methyl-terminated mesopore walls with neutral properties with respect to nonvolatile ink components. Hence, non-volatile and particulate ink components are not attracted by the mesopore walls – in contrast to porous oxidic stamps with hydroxyl-terminated pore walls or polymeric stamps with pore walls consisting of polar polymers or block copolymer components.[10a-e] This feature of the mesoporous silica stamps helps prevent mesopore clogging by adsorption of ink components to the mesopore walls. 3) The rigid contact elements of the mesoporous silica stamps are stable against deformation during substrate contacts. Hence, capillary stamping with mesoporous silica stamp is a robust process tolerant against variations of stamp-substrate contact times and the pressure exerted on the mesoporous silica stamps. The size and the shape of the stamped patterns are determined by a complex interplay of the size and the shape of the stamps' contact elements, the ink-substrate contact angles and the volatility of the ink solvents.[10f, 10g] Such as the examples discussed by Schmidt et al.[10f] and Philippi et al.,[10g] the model cases tested here involve self-limited pattern formation – i.e., mechanisms that prevent unlimited growth of the deposited patterns or uncontrolled ink spreading.

So far, mesoporous silica stamps have predominantly been used for capillary stamping of silanes and alkyl thiols forming chemisorbed monolayers.[10f, 10g] However, first experiments with fullerene solutions as inks[10f] suggested that mesoporous silica stamps may be used for parallel additive substrate patterning – including the generation of nanoparticle arrays on the substrates – by lithographic deposition of organic solutions. Here, we used mesoporous silica stamps obtained by molding MTMS-containing sol-gel precursor solutions for silica against silicon molds accessible by standard photolithography and wet-chemical pattern transfer. The mesoporous silica stamps thus obtained with mean mesopore diameters of 31 nm and 38 nm were patterned with contact elements having base diameters of ≈600 nm, tip diameters of ≈400 nm and heights of ≈500 nm forming trigonal arrays with a nearest neighbor distance of ≈1.3 µm (Supporting Figure S2).

## 3.2 Inks containing functional low-molecular mass compounds – 7α-ethinylestradiol (EE$_2$) nanoparticles

To exemplarily demonstrate capillary stamping of inks containing functional low-molecular mass compounds like drugs, we stamped a 50 mM solution of EE$_2$ in acetonitrile onto glass slides coated with the perfluorinated silane 1H,1H,2H,2H-perfluorodecyltrichlorosilane (FDTS). The FDTS coating reduced the adhesion of the formed EE$_2$ nanoparticles to the substrates and facilitated their transfer to aqueous colloidal nanodispersions. The positions of the stamped EE$_2$ nanoparticles imaged by scanning electron microscopy (SEM) (Figure 2a-c) did not constitute a perfect hexagonal lattice. Their initial arrangement reflecting the ordering of the stamps' contact elements was not conserved under the conditions of SEM imaging. The evaluation of 154 $EE_2$ nanoparticles identified in a SEM image yielded an average nanoparticle diameter of 123 nm ± 19 nm (Supporting Figure S3). AFM measurements were difficult to perform since the $EE_2$ nanoparticles were partially displaced by interactions with the cantilever tips. However, based on topographic AFM images we estimated the height of the $EE_2$ nanoparticles to ~11 nm (Supporting Figure S4). Thus, the $EE_2$ nanoparticles had a disk-like shape with larger specific surface and less pronounced tendency to agglomerate than spherical nanoparticles. To transfer the $EE_2$ nanoparticles to aqueous colloidal nanodispersions, we detached the $EE_2$ nanoparticles from the FDTS-coated glass slides by sonication (Figure 2d,e). In typical experiments, we stamped 10 $EE_2$ nanoparticle arrays next to each other and transferred the $EE_2$ nanoparticles then into 2 mL deionized water. Assuming that capillary stamping yielded defect-free hexagonal arrays of $EE_2$ nanoparticles with a lattice constant of 1.3 µm and that all $EE_2$ nanoparticles were transferred, the aqueous $EE_2$ nanoparticle dispersions obtained here would contain ≈170 million $EE_2$ nanoparticles corresponding to $EE_2$ concentrations of ≈0.02 nmol/mL or ≈7 ng/mL. The formation of drug nanoparticles and their transfer to colloidal nanodispersions can potentially be automated and optimized by adapting commercially available stamping devices designed for polymer pen lithography. For example, it is conceivable that the drug nanoparticles are stamped on a flexible foil mounted on a conveyor system that moves the foil, on the one hand, forward after each stamping step. On the other hand, the conveyor system would move the foil carrying the drug nanoparticles through a water bath where the drug nanoparticles are detached from the foil so that they enrich in the water bath until the target concentration is reached.

Further molecular building blocks can be coupled to the $EE_2$ molecules *via* their terminal triple bonds. We exploratively functionalized $EE_2$ nanoparticles dispersed in water by coupling the dye sulfo-cyanine-3-azide to exposed $EE_2$ triple bonds. Residual dye was then carefully removed from the reaction mixture by dialysis. Thus, the color of the reaction mixture containing the dye-modified $EE_2$ nanoparticles turned from pink to colorless. However, fluorescence emission spectroscopy revealed that the $EE_2$ nanoparticles were functionalized with sulfo-cyanine-3-azide (Figure 2f, black dotted curve). A band at ≈563 nm and a shoulder at ≈600 nm, which are characteristic fluorescence features of sulfo-cyanine-3-azide, were present both in the fluorescence emission spectrum of a colloidal dispersion of $EE_2$ nanoparticles reacted with sulfo-cyanine-3-azide and in a reference spectrum (dashed red curve in Figure 2f) obtained from a 1 mM aqueous sulfo-cyanine-3-azide solution. The featureless solid blue curve in Figure 2f is a fluorescence spectrum taken from a colloidal $EE_2$ nanoparticle dispersion prior to the reaction with sulfo-cyanine-3-azide. Dynamic light scattering (DLS) measurements on the colloidal $EE_2$ nanoparticle dispersions (Table S1, Supporting Information) revealed an apparent DLS diameter of ≈70 nm after storage times up to 1h, indicating the absence of larger $EE_2$ nanoparticle agglomerates. Progressing agglomeration after longer storage times, as indicated by increases in the apparent DLS diameters and the emergence of a second particle population with apparent DLS diameters of a few 100 nm, was reversible; the $EE_2$ nanoparticles could be redispersed nearly completely by sonication for 1 h. The precipitation obtained by deposition of a colloidal EE$_2$ nanoparticle dispersion onto a glass slide had a clear nanoparticulate character and consisted of disk-like EE$_2$ nanoparticles with diameters in the 100 nm range (Figure 2g).

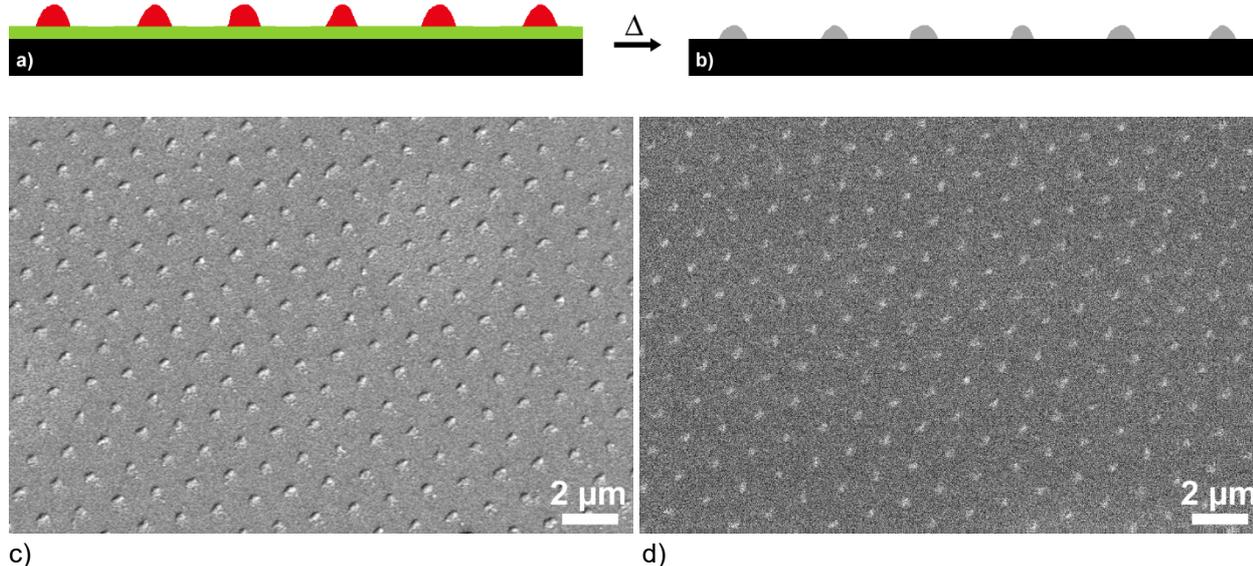

c)                                                                 d)

**Figure 3**. Pyrolytic conversion of PFS homopolymer nanoparticles obtained by capillary stamping to ceramic Fe/SiC nanoparticles. a), b) Schematic representation. a) PFS nanoparticles (red) stamped onto a quartz substrate (black) modified with FDTS (green); b) ceramic Fe/SiC nanoparticles (grey) form as the pyrolysis products of the PFS nanoparticles. c), d) SEM images of c) PFS nanoparticles stamped onto a FDTS-modified quartz substrate and d) ceramic nanoparticles obtained by pyrolysis of the PFS nanoparticles.

### 3.3. Inks containing homopolymers

Inks containing PFS homopolymer were tested as first model system for polymeric inks. We stamped a mixture of 25 vol-% toluene and 75 vol-% chloroform containing 2.5 mg/mL PFS onto FDTS-coated glass or quartz substrates (the latter were used for the pyrolysis experiments). The contact angle of the solvent mixture used here on FDTS-coated glass amounted to 55°. Stamping PFS homopolymer ink onto FDTS-coated quartz substrates yielded arrays of PFS nanoparticles (Figure 3a) that were subsequently converted to ceramic Fe/SiC nanoparticles by heating to 900°C at a rate of +40 to +41 K/min under argon (Figure 3b). The as-stamped PFS nanoparticles (Figure 3c) had diameters of 454 nm ± 52 nm, as determined by evaluation of 661 PFS nanoparticles imaged by SEM (Supporting Figure S5). Their heights amounted to 30 nm – 35 nm, as revealed by AFM (Supporting Figure S6). The ceramic Fe/SiC nanoparticles (Figure 3d) obtained as pyrolysis products had diameters of 221 nm ± 49 nm according to the evaluation of 231 ceramic Fe/SiC nanoparticles imaged by SEM (Supporting Figure S7) and heights reduced to 1-2 nm, as revealed by AFM (Supporting Figure S8). It is notable that, in contrast to the as-stamped EE$_2$ nanoparticles, the Fe/SiC nanoparticles still formed a well-ordered trigonal array reproducing the array ordering of the contact elements of the mesoporous silica stamps. We assume that the destruction of the FDTS layer in the course of the pyrolysis improves adhesion of the Fe/SiC nanoparticles to the quartz substrates.

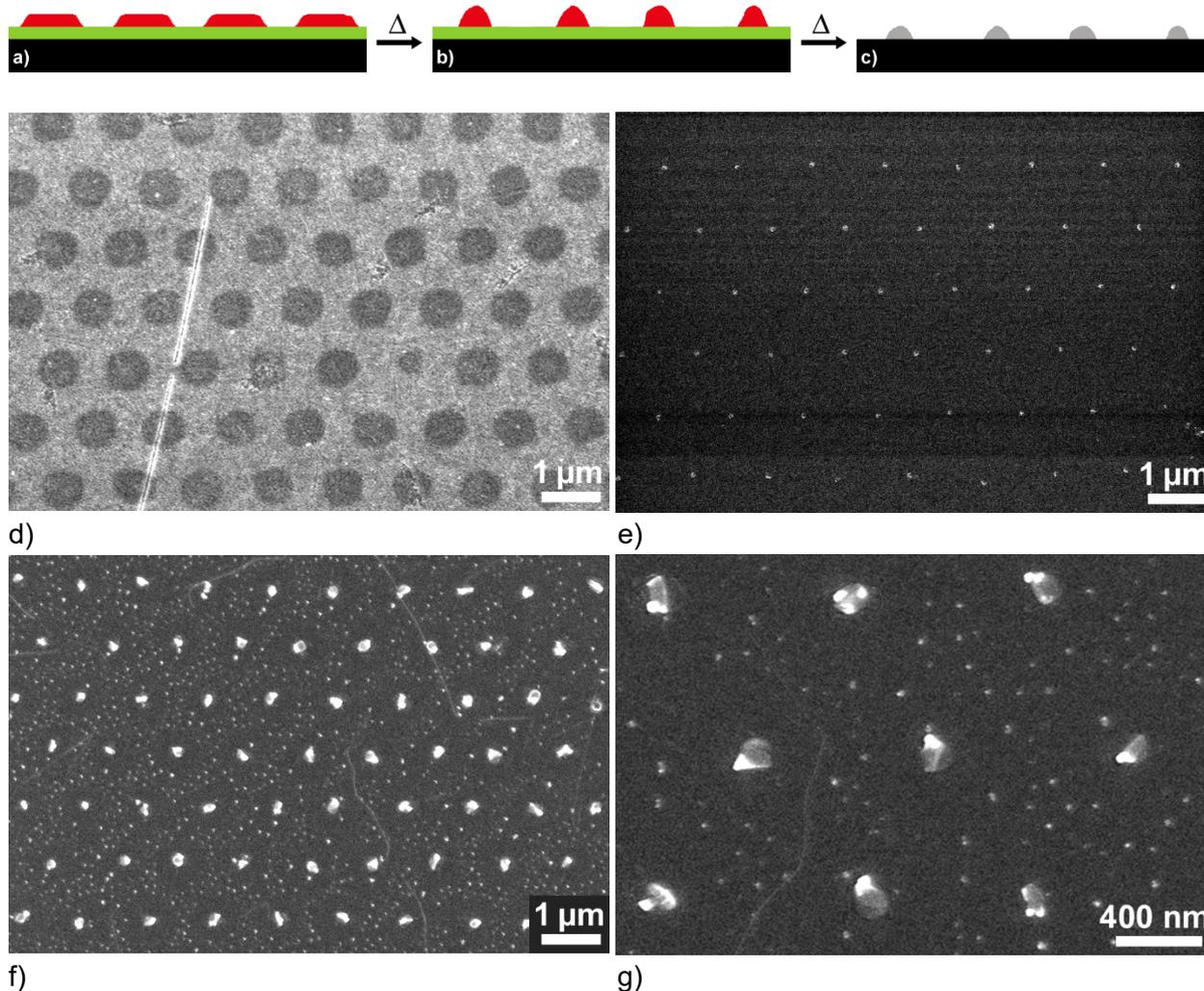

**Figure 4.** Flat PFS-*b*-P2VP microdots obtained by capillary stamping on FDTS-modified silica surfaces and Fe/SiC nanoparticles as their pyrolysis products. a) Flat PFS-*b*-P2VP microdots (red) are formed on a silica surface (black) coated with a FDTS layer (green). During the heating of the samples b) a temperature range is passed where the PFS-*b*-P2VP is softened and the flat PFS-*b*-P2VP microdots transform into 3D PFS-*b*-P2VP nanoparticles before c) pyrolysis at even higher temperatures converts the PFS-*b*-P2VP nanoparticles into Fe/SiC nanoparticles. d) SEM image of as-stamped flat PFS-*b*-P2VP microdots on a FDTS-modified glass slide; e) SEM image of a hexagonal array of Fe/SiC nanoparticles obtained by pyrolysis of flat PFS-*b*-P2VP microdots stamped on a FDTS-modified quartz surface. f), g) SEM images of carbon-ceramic hybrid nanoparticles obtained by pyrolysis of flat PFS-*b*-P2VP microdots stamped on a FDTS-modified quartz surface and subsequent high-temperature exposure to $H_2$/$CH_4$. f) Overview; g) detail.

### 3.4. Inks containing block copolymers

We stamped an ink containing 2.5 mg/mL PFS-*b*-P2VP in a mixture of 25 vol-% toluene and 75 vol-% chloroform onto FDTS-coated glass or quartz substrates. Thus, the inks used to stamp PFS-*b*-P2VP and PFS homopolymer were – apart from the polymeric component – identical. However, in striking contrast to PFS homopolymer, stamping ink containing PFS-*b*-P2VP onto FDTS-coated glass slides yielded arrays of flat PFS-*b*-P2VP microdots (Figure 4a). During pyrolysis, the flat microdots reorganize into 3D nanoparticles (Figure 4b) before they are converted into Fe/SiC nanoparticles (Figure 4c). SEM analysis revealed that the flat PFS-*b*-P2VP microdots formed ordered arrays reproducing the arrangement of the contact elements of the mesoporous silica stamps (Figure 4d, Supporting Figure S9). Analysis of 70 flat PFS-*b*-P2VP microdots imaged by SEM revealed a nearest-neighbor distance of ~1.3 µm and a mean diameter of 606 nm ± 42 nm (Supporting Figure S10). AFM imaging revealed a height of the flat PFS-*b*-P2VP microdots not exceeding 3 nm - 4 nm (Supporting Figure S11), which is one order of magnitude smaller than the height of as-stamped PFS homopolymer nanoparticles. The striking morphometric difference between as-stamped PFS homopolymer nanoparticles and flat PFS-*b*-P2VP microdots might be related to the incompatibility of PFS and P2VP. The PFS and P2VP block segments repel each other and preferentially interact with the ink solvents. Under the dynamic conditions in the liquid bridges, the formation of close-to-equilibrium PFS-*b*-P2VP structures with particulate micelle-like phenotypes is likely prevented. As more PFS-*b*-P2VP is drawn into the liquid bridges, the favored swollen state is maintained by surface diffusion of the of the PFS-*b*-P2VP molecules on the FDTS-coated substrate surface. This leaking from the liquid bridges may drag the microscopic ink-substrate contact line beyond the position being expected when the macroscopic ink contact angle is considered.

A glass transition temperature $T_g$ of 112°C for P2VP homopolymer forming 160 nm thick films on silicon nitride substrates was obtained by nanocalorimetry,[29] while $T_g$ of PFS lies in the range of 33°C.[30] Moreover, the PFS blocks are known to show low nucleation rates and slow crystal growth. While it is unlikely that the PSF blocks crystallize at all under the conditions of capillary microstamping, the actual melting temperatures of PFS homopolymer and of PFS blocks in PFS-*b*-PS were reported to lie well below 150°C.[31] Hence, a temperature window exists, in which the PFS-*b*-P2VP molecules soften. As a result, the PFS-*b*-P2VP dewetted and formed particulate structures closer to the melt equilibrium structure. Another effect likely driving the conversion of flat PFS-*b*-P2VP microdots into 3D nanoparticles is the disjoining pressure.[32] Further temperature increase resulted in pyrolytic conversion of the PFS-*b*-P2VP nanoparticles into ceramic Fe/SiC nanoparticles still forming ordered arrays (Figure 4e). Analysis of 108 Fe/SiC nanoparticles imaged by SEM revealed that their diameter decreased to 90 nm ± 37 nm (Supporting Figure S12). As revealed by AFM measurements (Supporting Figure S13), their height amounted to ≈5 nm. We treated the ceramic Fe/SiC nanoparticles obtained from PFS-*b*-P2VP with $H_2$/$CH_4$ carbon precursor mixture at 900°C for 10 min, conditions previously used to grow carbon nanotubes (CNTs).[33] Under these conditions, carbon is deposited at iron grains in the Fe/SiC nanoparticles.[22] The deposition of carbon increased the nanoparticle sizes and the contrast in SEM images (Figure 4f,g). Evaluation of 147 thus-treated Fe/SiC nanoparticles imaged by SEM revealed a mean nanoparticle diameter of 148 nm ± 22 nm (Supporting Figure S14), whereas AFM measurements revealed a nanoparticle height of 15-20 nm (Supporting Figure S15). On all Fe/SiC nanoparticles treated with $H_2$/$CH_4$ bright protrusions can be seen in the SEM images (Figure 4g), which we assume to be carbonaceous and attached to exposed Fe grains. The carbon-ceramic hybrid nanoparticles formed hexagonal lattices and are, therefore, still located at the positions where the contact elements of the mesoporous silica stamps contacted the substrate. However, smaller carbon-ceramic hybrid nanoparticles with diameters of a few 10 nm are also visible, which approximately mark the boundaries of Voronoi cells centering about the large carbon-ceramic hybrid nanoparticles defining the hexagonal lattice (Figure 4g,f and Supporting Figure S16). In between of the boundaries of the Voronoi cells populated by the small carbon-ceramic hybrid nanoparticles and the larger central carbon-ceramic hybrid nanoparticles depletion zones exist that contain hardly any nanoparticulate materials. Elongated structures crossing some of these zones are carbon nanotubes grown from small carbon-ceramic hybrid nanoparticles at the boundaries of the Voronoi cells.

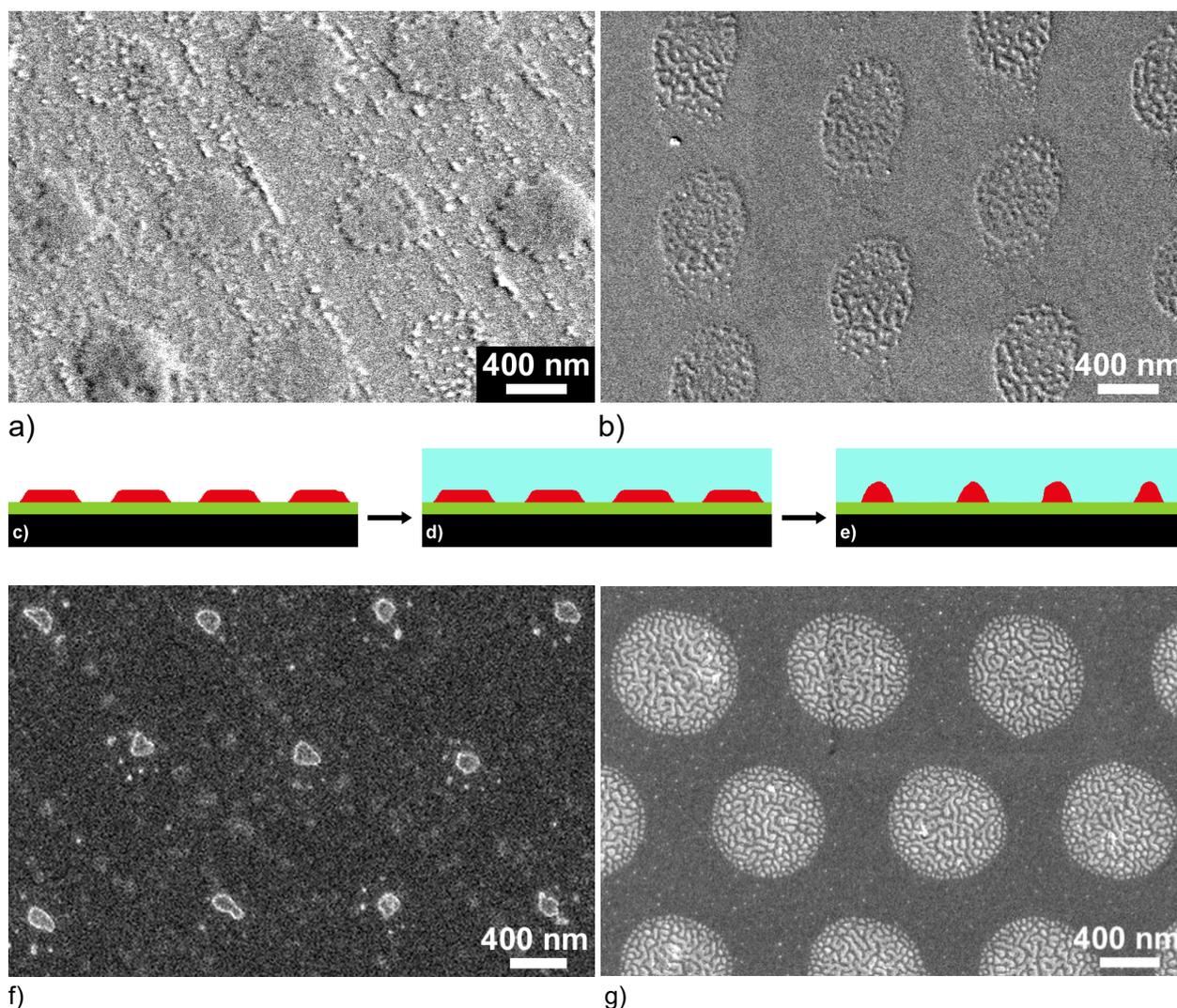

**Figure 5.** Capillary stamping of block copolymer inks on FDTS-coated and native hydroxyl-terminated substrates. a), b) Flat PFS-*b*-P2VP microdots stamped on a) FDTS-coated and b) uncoated hydroxyl-terminated glass substrates. c)-g) Conversion products of as-stamped flat PS-*b*-P2VP microdots obtained by treatment with ethanol at 60 °C for 1 h. c) – e) Scheme showing c) PS-*b*-P2VP microdots (red) deposited onto glass slides (black) coated with FDTS (green) by capillary stamping. d) Treatment with hot ethanol converts the PS-*b*-P2VP microdots into e) 3D PS-*b*-P2VP nanoparticles. f) SEM image of PS-*b*-P2VP nanoparticles on a FDTS-modified glass slide after treatment with ethanol at 60°C for 1h. g) SEM image of PS-*b*-P2VP microdots stamped on a native hydroxyl-terminated glass slide after treatment with ethanol at 60°C for 1h.

We rationalize this observation as follows. Upon heating, the PFS-*b*-P2VP molecules on the FDTS-coated substrate surface get mobile and undergo surface diffusion. It is straightforward to assume that, for geometrical reasons, the PFS-*b*-P2VP molecules agglomerate in the centers of the PFS-*b*-P2VP microdots. The number of PS-*b*-P2VP molecules in an annular ring segment of a PFS-*b*-P2VP microdot with radius $r_1$ is larger than that in an annular ring segment with radius $r_2$ and the same width if $r_1 > r_2$. The center of a PFS-*b*-P2VP microdot contains less PS-*b*-P2VP molecules than any annular ring segment surrounding the center. Therefore, more PFS-*b*-P2VP molecules will diffuse toward the center than away from the center. The agglomerates of PFS-*b*-P2VP molecules at the positions of the microdot centers thus evolve into 3D PFS-*b*-P2VP nanoparticles, which are stable because the PSF-*b*-P2VP molecules form microphase morphologies and adapt conformations closer to equilibrium than in the flat PFS-*b*-P2VP microdots. On the other hand, smaller agglomerates of PFS-*b*-P2VP molecules may form away from the positions of the microdot centers. However, the bigger 3D PFS-*b*-P2VP nanoparticles at the positions of the microdot centers will grow at the expense of the smaller agglomerates with larger surface-to-volume ratios because of the occurrence of Ostwald ripening. Only smaller PFS-*b*-P2VP aggregates far enough away from the positions of the microdot centers, that is, close to the Voronoi cell boundaries, survive. As a result, the patterns seen in Figure 4f and g are obtained after pyrolysis and exposure to the $H_2/CH_4$ carbon precursor mixture.

The nature of the substrate surface may influence post-stamping treatments for the morphometric modification of as-stamped patterns more severely than capillary stamping itself. Capillary stamping of PFS-*b*-P2VP solutions on FDTS-coated (Figure 5a) and uncoated hydroxyl-terminated (Figure 5b) glass substrates yielded morphometric similar PFS-*b*-P2VP microdots. Evaluation of 44 PFS-*b*-P2VP microdots stamped on a hydroxyl-terminated glass substrate imaged by SEM revealed a microdot diameter of 571 nm ± 41 nm (Supporting Figure S17) and a center-to-center distance of 1.3 µm. AFM revealed that the height of the PFS-*b*-P2VP microdots did not exceed 2 nm (Supporting Figure S18). However, after heating PFS-*b*-P2VP microdots stamped on uncoated hydroxyl-terminated quartz substrates to 600°C no pyrolysis products were found. P2VP blocks of BCPs containing a second nonpolar block such as PFS or PS are known to selectively segregate to oxidic surfaces,[23b, 34] where hydrogen bonds form between surface hydroxyl groups and the free electron pairs of the pyridyl nitrogen atoms of the P2VP repeat units. The hydrogen bonds prevent significant center-of-mass surface diffusion of BCP molecules containing P2VP block segments. Consequently, the reorganization of the PFS-*b*-P2VP microdots into 3D particulate structures – a prerequisite for the pyrolytic formation of ceramic nanoparticles – is prevented.

The impact of the substrate surface on post-stamping treatments can exemplarily be illustrated by exposure of PS-*b*-P2VP patterns stamped onto FDTS-coated and uncoated hydroxyl-terminated glass slides with hot ethanol. At first, capillary stamping was carried out with an ink containing 2.5 mg PS-*b*-P2VP per mL of a solvent mixture consisting of 25 vol-% toluene and 75 vol-% chloroform. Presumably, discrete PS-*b*-P2VP microdots with a smooth surface formed (Figure 5c) that resembled the PFS-*b*-P2VP microdots displayed in Figures 4d and 5a. However, we could hardly detect as-stamped PS-*b*-P2VP structures by SEM on either substrate type because PS-*b*-P2VP does not contain heavy atoms such as PFS-*b*-P2VP. On FDTS-modified glass slides, treatment with ethanol at 60°C for 1 h (Figure 5d) yielded 3D PS-*b*-P2VP nanoparticles (Figure 5e) arranged in regular arrays still retaining the ordering imposed by the mesoporous silica stamps (Figure 5f). Analysis of 183 3D PS-*b*-P2VP nanoparticles imaged by SEM revealed a nearest-neighbor distance of ~1.3 µm and a mean nanoparticle diameter of 191 nm ± 21 nm (Supporting Figure S19). The 3D PS-*b*-P2VP nanoparticles had heights up to ~60 nm, as revealed by AFM scans (Supporting Figure S20). We assume that, such as in the course of the pyrolysis of PFS-*b*-P2VP, enhanced molecular mobility is crucial for the morphometric reconstruction of the as-stamped BCP pattern. The PS-*b*-P2VP molecules did not desorb when exposed to hot ethanol because adsorbed polymers form multiple focal substrate contacts, the simultaneous loosening of which is highly unlikely. Nevertheless, exposure to hot ethanol facilitates center-of-mass displacements of individual PS-*b*-P2VP molecules by surface diffusion involving successive movements of chain segments. Thus, the PS-*b*-P2VP molecules assembled into 3D PS-*b*-P2VP nanoparticles, which are closer to the micellar PS-*b*-P2VP equilibrium structure in ethanol than PS-*b*-P2VP microdots. The AFM image seen in Supporting Figure S20a reveals that, besides the larger 3D PS-*b*-P2VP nanoparticles forming the stamped trigonal lattice, some smaller PS-*b*-P2VP nanoparticles with diameters in the 10 nm range are visible also. This observation indicates, such as for the thermal annealing of PFS-*b*-P2VP microdots, structure evolution dominated by Ostwald ripening.

On uncoated hydroxyl-terminated glass substrates treatment with ethanol at 60°C for 1 h yielded PS-*b*-P2VP microdots with porous-corrugated surface topographies that could easily be imaged by SEM (Figure 5g) and AFM (Supporting Figure S21). Evaluation of 89 PS-*b*-P2VP microdots imaged by SEM revealed an average center-to-center distance of 1.3 µm and a mean microdot diameter of 730 ± 24 nm (Supporting Figure S22). AFM topography analysis (Supporting Figure S21b) revealed a microdot height of 2 nm – 5 nm. The porous-corrugated surface topography originated from swelling-induced morphology reconstruction.[23a, 23c] Since hot ethanol is a good solvent for P2VP and a non-solvent for PS, swelling P2VP blocks migrated to the outer surfaces of the PS-*b*-P2VP microdots. When the ethanol evaporated, nanopores remained in place of the swollen and expanded P2VP block segments. Nevertheless, the PS-*b*-P2VP molecules remained pinned to their positions on the substrate surface; morphological changes were limited to the local scale defined by the molecular dimensions of the PS-*b*-P2VP chains. It is remarkable that capillary stamping yielded discrete BCP microdots on hydroxyl-terminated glass substrates because the ink solvents are expected to spread (macroscopic contact angles are 0°). The formation of the BCP microdots under these conditions obviously shows some parallels to capillary stamping of alkylthiol microdots onto gold surfaces.[10f] At the spreading front ink spreading competes with depletion of the ink solvents by evaporation followed by adsorption of the BCP molecules onto the substrate surface. At the positions of the stamp's contact elements dense microdot-like BCP layers eventually cover the substrate surface. Further ink drawn out of the contact elements senses the BCP layer rather than the oxidic substrate surface. On P2VP and PS homopolymers, which represent the block segments of the model BCP PS-*b*-P2VP, the ink solvent mixture has macroscopic contact angles of 32° and 34°. Hence, screening of substrate-solvent interactions by an adsorbed PS-*b*-P2VP layer in the course of capillary stamping likely contributes to the formation of discrete PS-*b*-P2VP microdots.

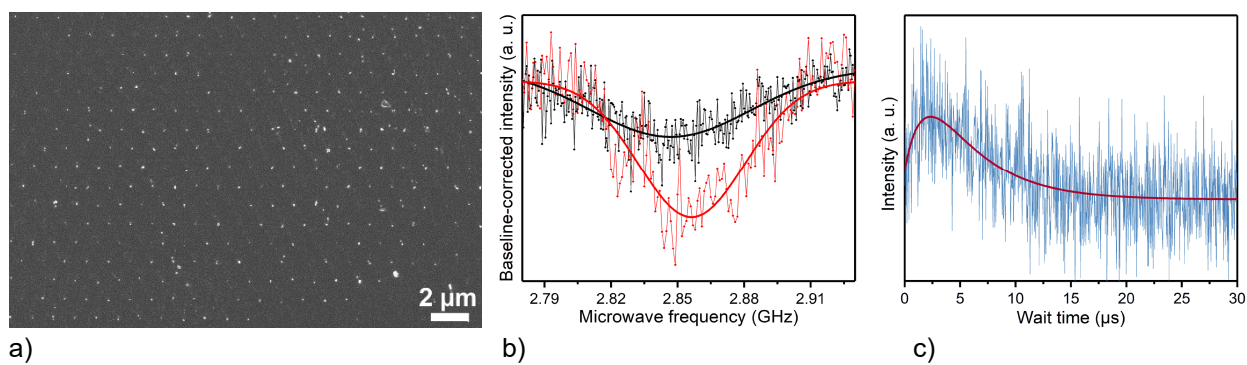

**Figure 6.** ND nanoaggregates deposited by capillary stamping of colloidal ND nanodispersions onto FDTS-coated glass slides. a) SEM image. b) ODMR measurements taken at a relative microwave power of -5 dB (black data points) and +7 dB (red data points). The lines connecting the data points are guides to the eyes. The solid black and red curves are Gaussian fits. c) $T_1$ relaxation time measurement. The red curve is a double exponential fit.[35]

### 3.5. Colloidal nanodispersions as inks: stamping of nanodiamond nanoaggregates.

Finally, we tested whether colloidal nanodispersions can be used as inks for capillary stamping, since in this way arrays of small aggregates of dispersed, preformed nanostructures may be accessible. Dispersions of 10 µg NDs with a nominal diameter of 10 nm per mL isopropanol were infiltrated into mesoporous silica stamps from their non-structured backsides. The mesoporous silica stamps acted as filters and rejected larger ND aggregates. Moreover, the confinement imposed by the mesopore walls prevented agglomeration of the NDs inside the mesoporous silica stamps. Finally, the relatively weak interactions between the NDs and the methyl-terminated mesopore walls reduced clogging. Thus, smaller NDs passed the mesoporous silica stamps and were deposited on FDTS-coated glass slides within the liquid bridges formed at the positions of the stamps' contact elements (the contact angle of isopropanol on FDTS-coated glass amounts to ≈45°). After retraction of the mesoporous silica stamps, hexagonal patterns of ND nanoaggregates were obtained (Figure 6a). Analysis of a SEM image with 334 ND nanoaggregates revealed a nearest-neighbor distance of 1.1 µm and an average ND nanoaggregate diameter of 43 nm ± 24 nm (Supporting Figure S23). AFM analysis was impeded by the poor adhesion of the ND nanoaggregates on the FDTS-coated glass surface; the AFM tip typically displaced the imaged ND nanoaggregates. From the AFM image shown in Supporting Figure S24 we nevertheless estimated the heights of the ND nanoaggregates to 30-40 nm. Hence, a ND nanoaggregate contained ≈10 NDs. The ND nanoaggregates obtained by capillary stamping were, therefore, somewhat smaller than previously reported ND nanostructures.[27] To prove the presence of stamped NDs on the substrate surface, we measured the characteristic response of diamond-hosted negatively charged nitrogen-vacancy (NV⁻) centers to standard optically detected magnetic resonance (ODMR) protocols.[36] The zero field splitting parameter $D$ and the longitudinal relaxation time $T_1$ are fingerprints for this defect.[37] $D$ is proportional to the energy difference between spin sublevels $|0\rangle$ and $|\pm1\rangle$ in the absence of external magnetic fields. A characteristic value for $D$ in nanodiamonds is 2.87 GHz for the electronic ground state.[38] The spontaneous transition rate between different sublevels is given by the $T_1$ relaxation time, which is typically in the microsecond range at room temperature for small nanodiamonds.[35] The observed continuous wave spectra (Figure 6b) were approximated by a single Gaussian model and revealed $D$ values of 2.856 GHz and 2.847 GHz for relative microwave power levels of -5 dB and +7 dB, respectively. Integration of the Gaussian fit revealed relative areas of 11.7 and 20.4 (in arbitrary units). The center frequencies mutually differ slightly due to thermal background contributions. The deviation from zero magnetic field resonance at room temperature may arise from crystal strain typically observed for small nanodiamonds.[38] The missing of a fine structure is likely caused by power broadening. Nevertheless, the measurements were carried out below the saturation limit and increasing the microwave power led to enhanced signal areas, as expected.[39] The $T_1$ measurement (Figure 6c) revealed spin relaxation times typical for small nanodiamonds.[35] Data fitting according to a double exponential model revealed a $T_1$ value of 1.8 μs. Since, according to the supplier's specifications, only ~6 % of the NDs contained NV⁻ centers, only a fraction of the stamped ND nanoaggregates contained NV⁻ centers so that direct imaging of the ND nanoaggregate arrays by detection of NV⁻ centers was not possible.

## 4. Conclusions

We have evaluated capillary stamping with mesoporous silica stamps as process platform for parallel additive substrate patterning and for the formation of nanoparticle arrays. So far, additive patterning has been conducted by serial ballistic methods associated with certain throughput limitations or by parallel microcontact printing with solid elastomeric stamps suffering from ink depletion after a limited number of stamp-substrate contacts. Mesoporous silica stamps enable continuous or triggered ink supply to their contact surfaces anytime during use *via* their mesopore systems. Thus, the problem of ink depletion is overcome while the throughput advantages of parallel stamping are maintained. Additive parallel capillary stamping with mesoporous silica stamps was tested using model inks containing low-molecular mass compounds such as drugs, homopolymers and block copolymers as well as colloidal nanoparticles. As model ink containing a low-molecular-mass drug, a solution of the contraceptive 17α-ethinylestradiol was stamped onto silanized glass slides. The drug nanoparticles obtained in this way were transferred to aqueous colloidal nanodispersions. Secondly, we tested polymeric inks and the post-stamping modification of the stamped polymeric patterns by pyrolysis or solvent treatments. Pyrolysis of stamped poly(ferrocenyldimethylsilane) nanoparticle arrays yielded arrays of ceramic Fe/SiC nanoparticles – a surface pattern, which is of interest for the guided growth of carbonaceous nanostructures. Capillary stamping of inks containing BCPs was also studied because of the BCPs' intrinsic capability to self-organize. As-stamped BCP patterns consisted of arrays of flat BCP microdots with a thickness a few nm. Only on silanized substrates dewetting induced by thermal annealing or solvent treatment resulted in the conversion of as-stamped flat BCP microdots into more compact 3D BCP nanoparticles with structures closer to equilibrium. This morphometric reconstruction appears to be the prerequisite for the pyrolytical conversion of flat PFS-*b*-P2VP microdots into Fe/SiC nanoparticles. As third model ink, we tested nanodiamond suspensions as example of colloidal inks containing dispersed preformed nanostructures and generated exemplarily arrays of nanodiamond nanoaggregates. The stamps' mesopore systems acted as filters rejecting large nanodiamond clusters and the confinement imposed by the mesopore walls prevented clustering of the nanodiamonds inside the mesoporous silica stamps. We anticipate that automatization of capillary stamping using commercially available stamping devices developed for polymer pen lithography enables the integration of parallel additive substrate manufacturing into continuous production processes. Functionalized substrates as well as nanoparticles – either as substrate-bound arrays or as colloidal nanodispersions after detachment – may be accessible with little consumption of solvents.

## 5. Experimental
### 5.1 Synthesis of PFS and PFS-*b*-P2VP

*Materials.* All solvents and reagents were purchased from Alfa Aesar, Sigma Aldrich, Fisher Scientific, ABCR or Merck and used as received unless otherwise stated. Tetrahydrofuran (THF) was distilled from sodium/benzophenone under reduced pressure (cryo-transfer) prior to the addition of 1,1-diphenylethylene (DPE) and *n*-butyllithium (*n*-BuLi) followed by a second cryo-transfer. Dimethyl[1]silaferrocenophane was synthesized and purified as described elsewhere.[40] 2-Vinylypridine (2VP) was purified by 2-fold distillation over calcium hydride ($CaH_2$) followed by treatment with trioctyl aluminium (25 wt.% in hexane), which was added to the monomer dropwise until a pale yellow color appeared. Prior to use within anionic polymerization protocols, 2VP was freshly distilled from this solution. DPE was degassed and *sec*-BuLi was added dropwise until a red color appeared, followed by distillation under reduced pressure. Degassed 1,1-dimethyl-silacyclobutane (DMSB) was treated with 1,1-diphenylhexyllithium (prepared by mixing 1 eq. DPE and 1 eq. *sec*-BuLi) until a pale red color maintained, followed by distillation under reduced pressure. Deuterated solvents were purchased from Deutero GmbH, Kastellaun, Germany. All syntheses were carried out under an atmosphere of nitrogen using Schlenk techniques or a glovebox equipped with a Coldwell apparatus.

*Synthetic procedures.* PFS and PFS-*b*-P2VP were prepared in a similar way as reported previously.[16c] In an ampoule equipped with a stirring bar, 850 mg dimethyl[1]silaferrocenophane (3.51 mmol, 108 eq.) and 14 mg LiCl (0.34 mmol, 10 eq.) are dissolved in dry THF (30 mL). The polymerization is initiated via rapid addition of 21.3 µL *n*-BuLi solution (0.03 mmol, 1.6 M, 1 eq.). After stirring the solution at room temperature for 3 h, the reaction is quenched by the addition of degassed methanol in the case of PFS homopolymer formation. For BCP formation with 2VP, 24.0 µL DPE (0.14 mmol, 4 eq.) and 8.7 µL DMSB (0.07 mmol, 2 eq.) are added to the reaction mixture prepared as described above containing active PFS macroinitiators followed by stirring for an additional hour at room temperature. After 1 h, an aliquot of the solution is taken from the ampoule for characterization of the PFS homopolymer after terminating the active chains by adding degassed methanol. The reaction mixture containing the residual active PFS chains is cooled to -78 °C. A prechilled solution of 110 mg 2VP (1.05 mmol, 31 eq.) in 2 mL THF is added to the active and end-functionalized PFS chains followed by stirring for 1 h. After quenching the reaction with methanol, the solution is poured into a ten-fold excess of methanol for polymer precipitation. The polymer is filtered, washed with methanol and dried in vacuum.

*Characterization*. Standard size exclusion chromatography was performed with THF as the mobile phase (flow rate 1 mL min$^{-1}$) on a SDV column set from (SDV 1000, SDV 100000, SDV 1000000 obtained from Polymer Standard Service, Mainz) at 30 °C. Calibration was carried out using polystyrene standards. For PFS homopolymer and the PFS-*b*-P2VP used here the following results were obtained. PFS homopolymer: $M_n$ = 33500 g/mol; $M_w$ = 71900 g/mol; Đ = 2.15. PFS sample taken during BCP formation: $M_n$ = 30000 g/mol; $M_w$ = 31000 g/mol; Đ = 1.03, PFS-*b*-P2VP block copolymer: $M_n$ = 32000 g/mol; $M_w$ = 33000 g/mol; Đ = 1.03. $^1$H NMR spectra were recorded at 300 MHz on a Bruker DRX300 spectrometer (300 MHz, CDCl$_3$, 296 K): $\delta$ = 0.46 (br, 1-H); 1.98 (br, 4-H); 2.39 (br, 5-H); 4.01 (br, 3-H); 4.21 (br, 2-H); 6.39 (br, 8-H); 6.81 (br, 6-H); 7.12 (br, 9-H); 8.24 (br, 7-H) ppm.

## 5.2 Synthesis and characterization of mesoporous silica stamps

Mesoporous silica stamps with a mean pore diameter of 31 nm used for capillary stamping of nanodiamonds and EE$_2$ were prepared following procedures reported elsewhere.[10f] To synthesize mesoporous silica stamps for the stamping of PFS, PFS-*b*-P2VP and PS-*b*-P2VP, the procedure was modified in that 10 g of 10 mM aqueous acetic acid solution was mixed with 1 g Pluronic F127 as well as 0.5 g urea in an Erlenmeyer flask and 4.98 mL MTMS were added after vigorous stirring for 30 min at room temperature. Thus, mesoporous silica stamps with a mean pore diameter of 38 nm, a BET surface of 430 m$^2$/g and a specific pore volume of 2.2 mL/g were obtained (Supporting Figure S25). The nitrogen sorption measurements were performed with a Porotec Surfer device at 77 K on mesoporous silica stamps dried by supercritical drying as described elsewhere.[10f] Before any measurement, the samples were degassed at 250 °C for 10 h. To obtain the pore size distribution, the adsorption branch of the BET isotherm was evaluated by applying the method of Barrett, Joyner and Halenda (BJH).[41] The obtained data were analyzed with the program ASiQwin from Quantachrome Instruments. All mesoporous silica stamps extended 5 x 5 x 3 mm$^3$ and had ≈17 million contact elements. After their synthesis, the mesoporous silica stamps were kept in the wet state and stored in ethanol.

## 5.3 Substrates for capillary stamping

Glass slides with a thickness of 0.17 mm extending 18 × 18 mm$^2$ were obtained from VWR. The quartz substrates used for the stamping of PFS and PFS-*b*-P2VP were standard ST cuts purchased from Hoffman Materials. The glass slides and quartz substrates were subjected to oxygen plasma for 10 min at a pressure of 5 mbar for 10 min using a Diener femto plasma cleaner. For silanization with 1H,1H,2H,2H-perfluorodecyltrichlorosilane (FDTS; supplied by abcr GmbH), glass slides and quartz substrates were heated to 80 °C for 2 h in a sealed glass container in the presence of an excess of FDTS (≈2 μL). Prior to stamping, all silanized glass slides were washed with ethanol and dried with nitrogen. Non-silanized hydroxyl-terminated glass slides were used directly after the plasma treatment.

**5.4 Capillary stamping**

*General procedure*: As stamp holders, we used stainless steel cylinders with a height of 4.5 cm, a diameter of 2 cm, a mass of ≈27 g and a flat cylinder base. Using double-sided adhesive tape, elastomeric poly(dimethyl siloxane) (PDMS) films with a thickness of ~3 mm were glued onto the flat cylinder bases of the stamp holders. The elastomeric PDMS films were prepared using Sylgard 184 formulation (Dow Corning). Base and PDMS prepolymer were mixed at a ratio of 1:9, stirred for 10 min, and poured into a polyethylene mold. The mixture was then allowed to cure for 1 week at room temperature. After removing excess ink from the surfaces of the mesoporous silica stamps with tissue, the mesoporous silica stamps were glued onto the exposed surface of the PDMS films with double-sided tape. All capillary stamping experiments were carried out manually under ambient conditions with a stamp-substrate contact time of 5 s.

*Capillary stamping of $EE_2$.* $EE_2$ (17α-Ethinylestradiol) and acetonitrile (ACN) were obtained from Sigma-Aldrich. Prior to capillary stamping, the mesoporous silica stamps were immersed into ACN for 12 h and subsequently into a 50 mM solution of $EE_2$ in ACN for 2 h.

*Capillary stamping of PFS, PFS-b-P2VP and PS-b-P2VP.* Mesoporous silica stamps stored in ethanol were transferred in the wet state into a mixture of 25 vol-% toluene and 75 vol-% chloroform for 1 day. Prior to capillary stamping, the mesoporous silica stamps were immersed into the inks for 2 h. As inks, we used solutions containing 2.5 mg/mL polymer in a mixture of 25 vol-% toluene and 75 vol-% chloroform. Chloroform was obtained from Roth (Germany), toluene was obtained from Fisher (Germany). As polymeric components, we used PFS and PFS-*b*-P2VP synthesized as described above, as well as PS-*b*-P2VP ($M_n$(PS) = 27700 g/mol; $M_n$(P2VP) = 4300 g/mol, Đ = 1.04) obtained from Polymer Source (Canada).

*Capillary stamping of nanodiamonds.* Milled nanodiamonds (NDNV10nmMd10ml) with a nominal diameter of 10 nm were received from Adámas Nanotechnologies as aqueous colloidal nanodispersion containing 1 mg mL$^{-1}$ NDs. Prior to capillary stamping, the NDs were transferred to colloidal nanodispersions in isopropanol. For this purpose, 1 mL of the aqueous colloidal ND nanodispersion was centrifuged and washed with 1 mL water and two times with 1 mL isopropanol. The precipitation was redispersed in 100 mL isopropanol to obtain a colloidal nanodispersion containing 10 µg mL$^{-1}$ NDs to be used as ink. The ink was applied to the unstructured backsides of the initially dry mesoporous silica stamps until the latter were completely transparent.

**5.5. Post-stamping treatments**

*Preparation of aqueous $EE_2$ nanoparticle dispersions.* 10 $EE_2$ nanoparticle arrays extending 5 x 5 mm$^2$ were stamped next to each other onto FDTS-coated glass slides. The samples were then sonicated in 2 mL deionized water with a Sonorex RK 102 H sonicator (Bandelin) at a frequency of 35 kHz for 2 h at 50 °C.

*Estimation of the $EE_2$ concentration in the $EE_2$ nanoparticle dispersions*. We assume that the $EE_2$ nanoparticles generated by capillary stamping constitute a defect-free trigonal lattice with a lattice constant of 1.3 µm. The area *A* of a hexagon with an edge length of

$a$ = 1.3 µm corresponding to the lattice constant of the EE$_2$ nanoparticle array can be calculated as:

$$A = \frac{3}{2} \cdot a^2 \cdot \sqrt{3} = 4.39 \text{ µm}^2$$

The hexagon contains one EE$_2$ nanoparticle in its center. Moreover, one EE$_2$ nanoparticle is located at each of the six hexagon corners. Each of these 6 EE$_2$ nanoparticles touches overall three hexagons so that 1/3 EE$_2$ nanoparticle is counted for the considered hexagon. Therefore, the number $P$ of EE$_2$ nanoparticles that can be ascribed to a hexagon can be calculated as:

$$P = 1 + 6 \cdot \frac{1}{3} = 3$$

The total area $A_{tot}$ of the 10 successively stamped EE$_2$ nanoparticle arrays having areas of 5 x 5 mm$^2$ amounts to 250000000 µm$^2$, and the total number of stamped EE$_2$ nanoparticles $P_{tot} = A_{tot}/A$ to ≈170 million. We modelled the stamped EE$_2$ nanoparticles as spherical segments with a height $h$ of 11 nm and a segmental radius $s$ of 62 nm. The spherical radius $r$ amounting to 175 nm can be obtained from the following relation:

$$s = \sqrt{h \cdot (2r - h)}$$

The volume $V$ of a spherical segment can be calculated as follows:

$$V = \frac{\pi}{3} \cdot h^2 \cdot (3r - h) = 65,1 \text{ nm}^3$$

Therefore, the integrated volume of all stamped EE$_2$ nanoparticles amounts to $V_{tot}$ =1.11 • 10$^{13}$ nm$^3$. Assuming an EE$_2$ density of 1.21 g cm$^{-2}$,[42] the total mass of EE$_2$ within the dispersion is 0.013 µg.

*Click-reaction of suspended EE$_2$ nanoparticles with sulfo-cyanine-3-azide.* Sulfo-cyanine-3-azide was obtained from Lumiprobe (Germany). All steps were carried out at room temperature (≈22 °C). 5 µL of an aqueous 1 mM sulfo-cyanine-3-azide solution were added to 2 mL of an aqueous dispersion of EE$_2$ nanoparticles prepared as described above. The solution was slowly stirred for 2 h. The reaction solution was then washed in a dialysis tube (Servapor; pore diameter ≈2.5 nm; molecular weight cut-off 12000-14000; obtained from SERVA Electrophoresis GmbH) with 1 L deionized water for 5 days and then with 1 L of fresh deionized water for one more day.

*Pyrolysis of PFS and PFS-b-P2VP.* FDTS-functionalized quartz substrates stamped with PFS nanoparticles or PFS-*b*-P2VP microdots were pyrolized in a tube furnace under a constant flow of argon (1.5 sccm, quality 4.6) by heating at a rate of +40 to +41 K/min within 22 min to 900°C. Subsequently, the furnace was switched off, and the samples cooled to room temperature at the natural cooling rate of the furnace. The samples shown in Figure 4f and g were, after a temperature of 900°C had been reached, subjected to a mixture of CH$_4$ (quality 2.5, flow 0.52 sccm) and H$_2$ (quality 5.0, flow 0.70 sccm) for 10 minutes.

## 5.6 Characterization

SEM imaging was carried out with a Zeiss Auriga system applying an accelerating voltage of 3 keV. Images were acquired with a secondary electron chamber detector or with an in-lens detector. Prior to SEM analysis, the samples were coated with platinum/iridium alloy using a K575X Sputter Coater from Emitech. Analysis of SEM images was carried out with the software ImageJ. AFM measurements were carried out with a NT-MDT Ntegra device using gold-coated silicon cantilevers with a force constant of 0.003-0.13 N/m in the contact mode, except the AFM measurements on PFM homopolymer samples, which were carried out using aluminum-coated silicon cantilevers with a force constant of 30-70 N/m in the non-contact mode. Dynamic light scattering (DLS) measurements on aqueous colloidal $EE_2$ nanodispersions were carried out with a Zetasizer Nano Series (Malvern) at 25 °C applying a laser wavelength of 633 nm. Equilibration times amounted to 2 min and 15 single measuring cycles were performed for one measurement. Fluorescence emission spectroscopy was carried out with a Cary Eclipse Fluorescence Spectrometer from Agilent Technologies with the specified excitation wavelengths. Contact angle (CA) measurements were carried out in the sessile drop mode at 22 °C and a humidity of 37 % using a drop shape analyzer Krüss DSA100. To measure contact angles of mixtures of 25 vol-% toluene and 75 vol-% chloroform on PS and P2VP, we prepared PS ($M_w$ = 239000 g/mol, $M_n$ = 233000 g/mol, Đ = 1.03; Polymer Standards Service, Mainz) homopolymer films as well as P2VP ($M_w$ = 37500 g/mol, $M_n$ = 35000 g/mol, Đ = 1.07; Sigma-Aldrich) homopolymer films. To this end, 2.5 mg homopolymer was dissolved in 500 µL chloroform and drop-cast onto glass slides.

## 5.7 ODMR measurements on stamped ND arrays

ODMR measurements were conducted using a home-built confocal microscope. A 520 nm digital modulated diode laser (Swabian Instruments) was used for optical excitation in the constant wave mode and for initialization as well as readout in the pulsed mode. For the focusing of the laser beam and for the collection of the fluorescent light of $NV^-$ centers an oil immersion objective (PLAPON 60XO; NA = 1.42; Olympus) was used. The emitted photons were guided through a 25 µm pinhole, detected with an avalanche photodiode (SPCM-AQRH-24; Excelitas) and counted either by a NI PCI-6236 counter card (National Instruments) or by a FPGA-based counter (Swabian Instruments). The microwave was applied by means of a 30 µm copper wire fed from a MW generator (APSIN 6010; AnaPico) that was amplified by ~40 dB with a high-power amplifier (AS0104-30/17; Milmega). Experimental control was accomplished using an adapted version of the Python-based software suite Qudi.[29] The following pulse scheme was used for $T_1$ relaxation measurements: 3 µs laser pulse – 2 µs waiting/deshelving time – incrementation of the spin evolution time. To obtain a sufficient signal-to-noise ratio, ~$10^5$ repetitions were recorded.

## Supporting Information

Supporting Information is available from the Wiley Online Library or from the author.

**Acknowledgements**

M. R., M. P. and M. S. thank the European Research Council (ERC-CoG-2014, Project 646742 INCANA) for funding. M.G. thanks the DFG for partial financial support (GA2169/7-1). W.H. thanks the DFG for financial support (HA2169/7-1).

**Notes**

The authors declare no competing financial interest.

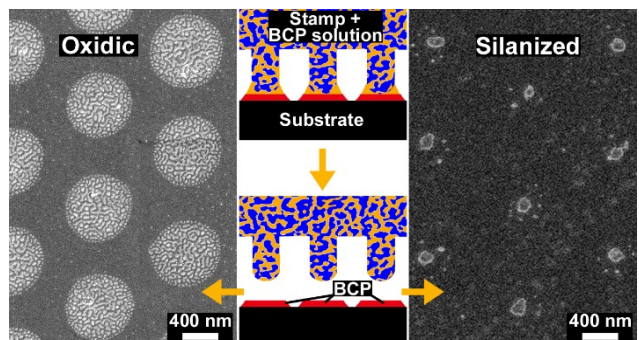
TOC Figure

TOC text: Capillary stamping with mesoporous silica stamps was explored as additive substrate patterning technique overcoming ink depletion, which limits the use of solid elastomeric stamps in state-of-the-art microcontact printing. Using selected model inks containing low-molecular-mass drugs, functional polymers or nanoparticles, nanoparticle arrays and nanodispersions of detached nanoparticles were obtained.